%% file: NNS.tex
\documentclass[conference]{IEEEconf}

\usepackage{balance} 

\input epsf
\usepackage{graphicx}
\usepackage{multirow}
\usepackage{titlesec}
\usepackage{balance} 
\usepackage{stfloats}
\usepackage{xcolor}

\usepackage{seqsplit}
\usepackage{listings}
\usepackage{xcolor}
\definecolor{mypurple}{RGB}{150,50,150}
\usepackage{subcaption}
\usepackage{subfig}
\usepackage{lineno}
\usepackage{tikz}
\usepackage{mathtools}
\usepackage{tabularray}
\usepackage{tabularx,calc,array}
\usepackage{amssymb}
\usepackage{float}
\usepackage{amsfonts}
\usepackage{booktabs}
\usepackage{multirow}
\usepackage{xurl}
\usepackage{comment}

\usepackage{amsmath}
\usepackage{verbatim}

\renewcommand\thesection{\arabic{section}} 

\renewcommand\thesubsectiondis{\thesection.\arabic{subsection}}

\renewcommand\thesubsubsectiondis{\thesubsectiondis.\arabic{subsubsection}}

\begin{document}

\title{\textbf{\Large The Economics of Blockchain Governance:\\
Evaluate Liquid Democracy on the Internet Computer\\}}
\author{Yulin Liu$^{l,*}$, and Luyao Zhang$^{z,*}$\\
 	\normalsize $^{l}$ SciEcon CIC, London, United Kingdom\\
  	\normalsize $^{z}$ Data Science Research Center, Duke Kunshan University, Suzhou, China\\
	\normalsize *L.Z. (lz183@duke.edu) and Y.L. (yulin.liu@sciecon.org) are the joint first and corresponding authors.
}


\maketitle
\begin{abstract}
Decentralized Autonomous Organizations (DAOs), utilizing blockchain technology to enable collective governance, are a promising innovation. This research addresses the ongoing query in blockchain governance: How can DAOs optimize human cooperation? Focusing on the Network Nervous System (NNS), a comprehensive on-chain governance framework underpinned by the Internet Computer Protocol (ICP) and liquid democracy principles, we employ theoretical abstraction and simulations to evaluate its potential impact on cooperation and economic growth within DAOs. Our findings emphasize the significance of the NNS's staking mechanism, particularly the reward multiplier, in aligning individual short-term interests with the DAO's long-term prosperity. This study contributes to the understanding and effective design of blockchain-based governance systems.
\end{abstract}
\IEEEoverridecommandlockouts
\vspace{1.5ex}
\begin{keywords}
\itshape Decentralized Autonomous Governance (DAO), Internet Computer Protocol (ICP), Liquid Democracy, Agent-based Modeling, Cooperation, Blockchain
\end{keywords}

%
\IEEEpeerreviewmaketitle

\section{Introduction}

A Decentralized Autonomous Organization (DAO) represents a groundbreaking approach to governance, utilizing blockchain technology to facilitate collaboration and enforce collective rules~\cite{ding2023survey}. Valerie's 2023 study~\cite{laturnus2023economics} on 2,377 Ethereum blockchain-based decentralized applications uncovers a correlation between on-chain governance activities and DAO valuations. DAOs with higher levels of voting participation tend to exhibit higher valuations, although the extent of decentralization does not significantly influence their valuation. Feichtinger et al. (2023)\cite{feichtinger2023hidden} identify challenges in the on-chain governance systems of 21 Ethereum-based DAOs, highlighting issues like concentrated voting rights, significant hidden costs, and a substantial volume of redundant governance activities. Bakos and Halaburda (2023)\cite{bakos2022will} discuss the tendency toward ownership concentration in DAOs when governance tokens are tradeable. Austgen (2023)~\cite{austgen2023dao} points out the risk of systemic bribery in DAOs, noting that this risk persists despite increases in decentralization. These findings emphasize the need for more research into the economic aspects of blockchain governance to design DAOs that are decentralized, secure, and efficient.

Our study introduces the Network Nervous System (NNS)—the most extensive on-chain governance mechanism, embodying the principles of liquid democracy on the Internet Computer Protocol (ICP)~\cite{liu2022pattern, hanke2018dfinity, dfinity2022internet, InnovateInternetComputer}. We examine the NNS’s potential to enhance human cooperation and economic growth within DAOs through theoretical abstraction and simulations.

The research aims to answer the following questions:
\begin{itemize}
\item How does the NNS structure its reward mechanisms to balance individual and collective interests, thereby promoting cooperation in blockchain governance?
\item What is the impact of the NNS mechanism on essential properties such as decentralization, security, and efficiency, and what strategies can improve its design?
\end{itemize}

Our findings suggest that the key to the NNS's effectiveness lies in its staking mechanism, particularly in the reward multipliers that align individual short-term interests with the long-term prosperity of the DAO. Section~\ref{sec: mechanism} abstracts and simulates the key features of the NNS mechanism. Section~\ref{sec:economics} demonstrates how its innovative design can improve decentralization, security, and efficiency through agent-based modeling simulations. The replicable code from our analysis is available on GitHub at~\url{https://github.com/SciEcon/nns}.  We invite readers to explore the interactive forms we’ve set up in Google Colab. These tools allow users to experiment with various parameters, such as duration, age, and participation, to observe their impact on voting power and rewards.
\section{The Mechanisms of NNS}
\label{sec: mechanism}
\input{figs/fig0}
 Figure~\ref{fig: nns} displays the NNS’s voting and reward design. 
Users can participate in the governance of ICP using NNS dApp.\footnote{https://github.com/dfinity/nns-dapp} Decisions on the Internet Computer are made by all stakeholders, marking ICP as a self-governing blockchain. Voting accounts on ICP are called “neurons”.  Users stake ICP tokens in neurons to gain voting power and rewards. The NNS incorporates two principal mechanisms to engage and to reward user participation in blockchain governance.

\subsection{The multipliers of neuron stake}
In NNS, the voting power and rewards are tied to the neuron stake, with a multiplier based on duration (dissolve delay) and age. The neuron stake is the number of staked ICP tokens. 

Figure~\ref{fig: multiplier_duration} highlights the dissolve delay’s impact on the multiplier, ranging from 0 to 8 years. Neurons with a dissolve delay under six months have no voting power. This six-month limit deters short-term voting, promoting long-term cooperation. Neurons\footnote{The statistics of neuron data can be queried from \url{https://dashboard.internetcomputer.org/neurons}.
} with longer duration and more tokens staked, get more voting power and rewards, with a duration of eight years providing the maximum multiplier of 2 times. Tokens can be retrieved by dissolving the neuron and become liquid when the delay reaches zero.
To motivate token holders to stay engaged, neurons that have been in a non-dissolving state for extended periods (referred to as “age”) benefit from increased multipliers on their voting power and rewards. Figure~\ref{fig: multiplier_age} showcases how age influences the multiplier, with a four-year duration yielding a peak multiplier of 1.25 times. 
\input{figs/fig2}
\input{figs/fig3}
\subsection{The staking reward}
A neuron’s voting reward is influenced by its action and the daily cumulative voting rewards. The neuron action pertains to governance participation activities, encompassing both direct voting and delegation. The daily voting reward represents the freshly minted ICP, calculated using a set inflation rate in Equation~\ref{eq:reward_rate} applied to the overall ICP supply. Specifically, let us denote the total supply of ICP at year \( y = \left\lfloor \frac{t}{12} \right\rfloor \)\footnote{\( \left\lfloor \frac{t}{12} \right\rfloor \) gives you the integer number of years from a total number of months \( t \), discarding any remainder.} by \( \text{I}_{y} \). The total reward at year \( y \), denoted by \( R_{y} \), is given by the product of the inflation rate \( i_{y} \) and total ICP supply \( \text{I}_{y} \):

\begin{equation}
R_{y} = i_{y} \cdot \text{I}_{y}, 
\end{equation}

\begin{equation}
I_{y+1} = (1+i_y) \times I_{y}.
\end{equation}
The yearly inflation rate \( i_{y} \) is defined as follows:

\begin{equation}
i_{y} = 
\begin{cases} 
5\% + 5\% \times \left(\frac{8 - y}{8}\right)^2 & \text{for } y \leq 8, \\
5\% & \text{for } y > 8.
\end{cases}
\label{eq:reward_rate}
\end{equation}

Given a total yearly reward \(R_{y}\), the basic idea is to distribute this reward evenly across each month of the year. This distribution is based on the assumption that the reward accrues consistently over time. Therefore, the initial calculation of the monthly reward \(R_{t}\) is simply a division of the yearly reward by the number of months in a year:

\[ R_{\text{t, base}} = \frac{R_{\text{y}}}{12} \]

By simulating the process that repeated annually over an eight-year period, starting with an initial supply of $469$ million ICP,  Figure~\ref{fig:inflation} illustrates the annual inflation rate and current supplies, while Figure~\ref{fig:ICP_reward} displays monthly and yearly reward distributions.

\input{figs/fig4}
\input{figs/fig5}

Let \( s_{it} \) be the stake of person \( i \) at time \( t \), \( a_{it} \) be the age multiplier for person \( i \) at time \( t \), and \( d_{it} \) be the dissolve delay multiplier for person \( i \) at time \( t \). 

Only \textit{Governors} who participate in NNS governance can claim rewards. Governors are defined as agents in either staking or dissolving status with a minimum of a 6-month dissolve delay before their neurons become liquid. The reward proportion for person \( i \) at month \( t \), \( p_{it} \), is given by Equation~\ref{eq:reward_proportion}:
\begin{equation}
\label{eq:reward_proportion}
p_{it} = \frac{s_{it} \cdot a_{it} \cdot d_{it}}{\sum_{j \in \text{Governors}} s_{jt} \cdot a_{jt} \cdot d_{jt}}.
\end{equation}

The final reward for person \( i \) at time \( t \), denoted by \( R_{it} \), is then calculated as:

\begin{equation}
\label{eq:final_reward}
R_{it} = R_{t} \cdot p_{it},
\end{equation}
where \( R_{t} \) is the total reward available at period \( t \). Then the monthly reward ratio, denoted by $r_{it}$ is:

\begin{equation}
\label{eq:final_reward_ratio}
r_{it} = \frac{R_{it}}{s_{it}} = \frac{R_{t}\cdot a_{it}\cdot d_{it}}{\sum_{j \in \text{Governors}} s_{jt} \cdot a_{jt} \cdot d_{jt}}
\end{equation}

If \( i \) is a governor, then \( r_{it} \) corresponds to the actual monthly reward ratio that the governor is entitled to claim, where the annualized reward ratio \( r^{annualized}_{it} \) is:
\begin{equation}
\label{eq:annualized_reward_ratio}
r^{annualized}_{it} = \frac{12 R_{it}}{s_{it}} = \frac{12 R_{t}\cdot a_{it}\cdot d_{it}}{\sum_{j \in \text{Governors}} s_{jt} \cdot a_{jt} \cdot d_{jt}}
\end{equation}

Users can visit the Internet Computer Governance Dashboard\footnote{\url{https://dashboard.internetcomputer.org/governance}} to evaluate the estimated reward ratio. This estimation takes into account their specified dissolve delay or liquidity preferences, which yields the annualized return for a neuron that is presumed to have an age of zero. To obtain this, one may annualize the reward ratio delineated in Equation~\ref{eq:final_reward_ratio} by setting \( a_{it} = 1 \):

\begin{equation}
\label{eq:final_reward_ratio_age0}
r^{\textit{estimated\_annualized}}_{it} = \frac{12R_{it}}{s_{it}} = \frac{12R_{t}\cdot d_{it}}{\sum_{j \in \text{Governors}} s_{jt} \cdot a_{jt} \cdot d_{jt}}
\end{equation}

In addition to voting, neurons can submit proposals. These proposals encompass a broad spectrum of topics, from expanding network capacity by integrating new nodes into the blockchain consensus to adjusting economic parameters like exchange rates and node provider rewards. To be eligible to submit a proposal, each neuron must hold a minimum amount of staked ICP tokens. This threshold typically aligns with the cost incurred if a majority vote doesn't ratify the proposal. Notably, this threshold is dynamic and can be proposed for modification through the NNS.\footnote{There are two majority rules: 1) absolute majority: before the voting period ends, a proposal is adopted or rejected if an absolute majority (more than half of the total voting power, indicated by delimiter above) has voted Yes or No on the proposal, respectively; 2) simple majority: When the voting period ends, a proposal is adopted if a simple majority (more than half of the votes cast) has voted Yes and those votes constitute at least 3\% of the total voting power. Otherwise, the proposal is rejected.} 

\section{The Economics of NNS}
\label{sec:economics}
This section delves into the economic principles that form the foundation of the NNS mechanisms, which are pivotal in fostering human collaboration and advancing economic prosperity within DAOs. We provide a detailed examination of the NNS economics through comparative analyses of agent-based modeling simulations.

\subsection{Initialize Agent Features}
We first initialize agents with a simulation framework. Each agent is characterized by three features: \textbf{endowment}, \textbf{staking threshold}, and \textbf{liquidity preference}. These features are key determinants of the agent's behavior within the model and are derived from specific statistical distributions to reflect variability and realism. 

\begin{itemize}
    \item \protect{\textbf{Endowment}}: The endowment of an agent is a measure of the resources allocated to it at the start of the simulation. It is sampled from a log-normal distribution, which is appropriate for modeling a wide range of natural phenomena, especially financial data. The distribution is parameterized by a mean (\texttt{mean\_log}) and a standard deviation (\texttt{sigma\_log}), both of which can be customized according to the desired characteristics of the agent population. The log-normal distribution is widely used for the wealth distribution, therefore for the token endowment distribution among agents simulated in this paper.
    \item \protect{\textbf{Staking Threshold}}: This feature determines the minimum level of estimated reward ratio that an agent is willing to commit to joining in the staking process. It is drawn from a gamma distribution, which is commonly used to model wait times or the amount of time until an event occurs. The shape (\texttt{k}) and scale (\texttt{theta}) parameters of the gamma distribution can be tuned to control the staking behavior of agents. Gamma distribution is widely used for waiting time. The staking threshold reflects the time agents need to wait until the staking return exceeds their staking threshold. 
    \item \protect{\textbf{Liquidity Preference}}: The liquidity preference reflects an agent's inclination towards maintaining liquid assets as opposed to staking them in the on-chain governance process. This is modeled using a custom bimodal distribution, allowing for the simulation of agents with two distinct types of financial behavior. The distribution is generated by a mix of two normal distributions with their own means (\texttt{mean1}, \texttt{mean2}) and a shared standard deviation (\texttt{std\_dev}). These parameters can be adjusted to shape the dual nature of the agents' liquidity preferences, which would be the agent's choice of dissolve delay if choose to stake. This distribution captures the liquidity preference of the two types of agents. Those with a strong liquidity preference would choose dissolving delay in the range of 6 months to 2 years, while those with a long-term vision choose to lock for 7-8 years to maximize the staking return. 
\end{itemize}

\subsection{Parameter Configuration}

\input{figs/fig6}
In the simulation, agents are initialized with designated initial endowments and staking thresholds based on preset distributions. To enhance the complexity and authenticity of each agent's profile, the liquidity preference is crafted using a CustomRandomVariableGenerator. This flexible structure allows for convenient modification and fine-tuning of agent attributes to meet the requirements of varied simulation settings. Figure~\ref{fig:agent_initialize} displays one initialization of agent features within this simulation framework based on the following parameter configurations.

\begin{itemize}
\item The number of agents $N = 10,000$. 
\item The parameters for agents' endowment $\texttt{mean\_log} = 10.57$ and $\texttt{sigma\_log}=0.6$.
\item The parameters for agents' staking threshold $\texttt{k}=1.8$ and $\texttt{theta}=0.055$.
\item the parameters for liquidity preference $\texttt{mean1}=18$, $\texttt{mean2}=96$, and $\texttt{std\_dev}=5$.
\end{itemize}

Under the configuration, the total endowment is around 469M, matching the initial supply of ICP in May 2021. And the mean of staking thresholds is around 10\%. 

\subsection{Simulate Agent Dynamics}
In the simulation, agent behavior is governed by the following decision-making criteria to act:

\begin{itemize}
\item \textbf{Macroeconomic Shocks}: The realized (adjusted) staking threshold of the agents depends on both the long-term staking threshold and the sentiment perturbations resulting from macroeconomic threshold shocks in each period.\footnote{For instance, a positive (negative) shock to crypto sentiments would result in a decrease (increase) in the staking threshold adjustments. That is, individuals would be inclined to stake at a lower (higher) anticipated reward following a positive (negative) sentiment shock arising from macroeconomic events. Examples of positive sentiment shocks include regulatory approvals and high-profile endorsements, while negative sentiment shocks may include regulatory crackdowns and security breaches.\begin{align*}
\texttt{adjusted\_staking\_threshold}^{i}_{t}= \\
\texttt{staking\_threshold}^{i} + \texttt{threshold\_shock}_{t}\\
\textit{Where:}\\
\texttt{threshold\_shock}_{t} = -\texttt{sentiment\_shock}_{t}
\end{align*}}

\item \protect{\textbf{Stake}}: Agents select to stake their entire token holdings when the estimated annualized reward ratio exceeds their adjusted individual staking threshold.
\item \protect{\textbf{Unstake}}: Conversely, agents opt to unstake their tokens if their realized annualized reward ratio falls below their adjusted staking threshold.

\end{itemize}

Otherwise, the agent will not act. These rules encapsulate the strategic decision-making process of the agents, balancing the potential rewards against their liquidity preferences. These rules encapsulate the strategic decision-making process of the agents, balancing the potential rewards against their liquidity preferences.

\subsubsection{The Benchmark}
In the benchmark scenario, the governance parameters are meticulously defined as follows:

\begin{itemize}
\item \protect{\textbf{Inflation Rate}}: This is maintained at a constant 5\% from the genesis point of the simulation.
\item \protect{\textbf{Dissolve Delay Multiplier}}: Set at 1.06, this parameter operates in conjunction with a dissolve delay period of six months.
\item \protect{\textbf{Age Multiplier}}: This parameter is held steady at a value of 1.
\end{itemize}
These settings provide a structured framework within which the dynamics of the simulation are evaluated, ensuring consistency and clarity in the governance rules applied throughout the benchmark analysis.

\subsubsection{The Comparative Studies}
In our comparative study, we assess the benchmark scenario alongside four distinct scenarios:
\begin{enumerate}
    \item \protect{\textbf{Scenario 1 - Inflation Rate Policy}}: Where only the inflation rate policy is adopted.
    \item \protect{\textbf{Scenario 2 - Dissolve Delay Multiplier Policy}}: Where only the dissolve delay multiplier policy is adopted.
    \item \protect{\textbf{Scenario 3 - Age Multiplier Policy}}: Where only the age multiplier policy is adopted.
    \item \protect{\textbf{Scenario 4 - Hybrid Policies}}: Where all three policies are adopted together.
\end{enumerate}

Figure~\ref{fig:gov_figures} displays the time series of governor and governance token counts. Figure~\ref{fig:token_pct_figures} illustrates the distribution of tokens in three states: liquid, staking, or dissolving in percentage. 

\begin{itemize}
    \item In comparison to the benchmark with a static 5\% inflation rate, \textbf{Scenario 1} implements a dynamic inflation rate that decreases from 10\% to 5\%. This change significantly reduces the proportion of liquid and dissolving tokens, while concurrently increasing both the number of governors and the quantity of tokens they stake throughout the simulation. The reasoning for this dynamic approach is based on the economic tenet that staked token holders receive increased rewards during periods of higher inflation, thus enhancing their share of the total token supply in contrast to that held by liquid token holders. Higher inflation rates, therefore, result in the dilution of liquid token holdings, making it less appealing to keep tokens in a liquid state. Nonetheless, it is crucial to highlight the evident trade-off with higher inflation rates, which involves the rapid expansion of the token supply surpassing that of the benchmark scenario. This necessitates the adoption of additional strategies in tandem with the dynamic inflation rate to curb the swell in token supply.
    \item In comparison to the benchmark, \textbf{Scenario 2}, characterized by a dissolve delay multiplier, displays a decrease in the percentage of liquid tokens alongside a marked increase in the percentage of dissolve tokens. This shift is driven by the reinforcing effect of the dissolve delay multiplier, where stakers with extended dissolve periods command a greater share of the staking reward due to higher accrued rewards for their patience. Although the total reward pool remains constant (as the inflation rate is consistent with the benchmark scenario), stakers with a strong liquidity preference incur a proportionally reduced reward, making them more susceptible to variations in staking returns. As a result, notable fluctuations are observed in the staking and unstaking activities of these stakers, who are sensitive to liquidity and have shorter dissolve delays.
    \item To alleviate volatility in staking ratios and reduce the proportion of dissolving tokens, focus shifts to \textbf{Scenario 3}, which introduces an age multiplier based on the benchmark. The results from the simulation indicate that the age multiplier has a moderate influence in decreasing the share of dissolving tokens. Stakers with shorter dissolve periods are able to counterbalance the dominant influence of those with extended dissolve delays by remaining in the ecosystem longer and accumulating neuron age. Nonetheless, the noticeable effect of this mechanism is relatively limited.
    \item \textbf{Scenario 4} integrates the three previously mentioned measures: a dynamic inflation rate, a dissolve delay multiplier, and an age multiplier. The outcomes demonstrate a significant decline in the proportion of liquid tokens and an increase in both the volume and the percentage of staked tokens.  Furthermore, the proportion of liquid tokens shows reduced volatility compared to the benchmark scenario.
\end{itemize}

In summary, the interplay of the dynamic inflation rate, age bonus, and dissolve delay mechanisms collectively contributes to the stabilization of the staking ratio's volatility, while concurrently sustaining a high level of staking or governance engagements.
\input{figs/fig8}
\input{figs/fig10}

\section{Conclusion and Future Research}
This paper explores the impact of dynamic inflation rates, age bonuses, and dissolve delays on the staking ratio. Future research could incorporate several key areas to enhance the comprehensiveness of the study:
\begin{itemize}
\item \textbf{Token Price Dynamics}: Extending the model to include token price dynamics \cite{10.1007/978-3-031-37717-4_51,9881799,10.1007/978-3-031-54053-0_39}\footnote{\url{https://coinmarketcap.com/currencies/internet-computer/}} could significantly enrich the study. Introducing an exogenous pricing factor, such as historical price data or predictive models for token price movements, could elucidate potential correlations or causal relationships between token price fluctuations and staking behaviors. This addition aims to provide a more comprehensive understanding of how token valuations influence staking decisions and the overall model dynamics.
\item \textbf{Market Sentiment and Behavioral Analysis}: Integrating market sentiment analysis \cite{10.1007/978-3-031-37963-5_7,Zhang_2023,quan2023decoding,fu2024dam} could offer insights into investors' perceptions and emotional responses to token developments. This understanding could deepen the analysis of how market sentiment directly influences staking decisions and contribute to a nuanced interpretation of staking behaviors.
\item \textbf{User Preferences and Risk Tolerance}: Employing behavioral economics methodologies, such as surveys or experiments, to assess user preferences and risk tolerances \cite{Xiao_2023, Zhang23,wu2024trust} could yield valuable insights. This approach could reveal how different risk profiles among stakeholders within the Web3 ecosystem affect staking ratios under varying conditions of inflation rates, age bonuses, and dissolve delays.
\item \textbf{Network Effects and Staking Adoption}: Analyzing the role of network effects \cite{ao2023decentralized, 10.1007/978-3-031-37717-4_67,yan2023validator,chemaya2023uniswap,augusto2023sok} in staking adoption can reveal how growth and broader adoption of a blockchain network influence staking behaviors. Investigating factors such as an increase in network participant inclusion, decentralization, and interoperability could provide rich insights into the dynamics of staking participation.
\item \textbf{Policy Implications and Economic Welfare}: Extending the research to evaluate the broader economic implications of staking mechanisms can assess their impact on economic welfare, market stability, and token distribution. Modeling different policy interventions or incentive structures \cite{zhang2023understand,zhang2023sok,10.1145/3548606.3559341} could help optimize staking participation while ensuring economic stability within the Web3 ecosystem.
\end{itemize}
\vspace{4mm}
By incorporating these future study directions, the research can delve deeper into understanding the intricate relationships between token dynamics, stakeholder behaviors, and the broader economic implications within the Web3 environment.

\section*{Acknowledgment}
Luyao Zhang is supported by the National Science Foundation China (NSFC) on the project entitled “Trust Mechanism Design on Blockchain: An Interdisciplinary Approach of Game Theory, Reinforcement Learning, and Human-AI Interactions (Grant No. 12201266). Luyao Zhang is also with SciEcon CIC, a not-for-profit organization based in the United Kingdom, aiming to cultivate interdisciplinary research of profound insights and practical impacts.

\bibliographystyle{ieeetr}
\bibliography{references}



\end{document}

%% file: figs/fig0.tex
\begin{figure}[!htbp]
\centering
\captionsetup{justification=centering}
\includegraphics[width=0.49\textwidth]{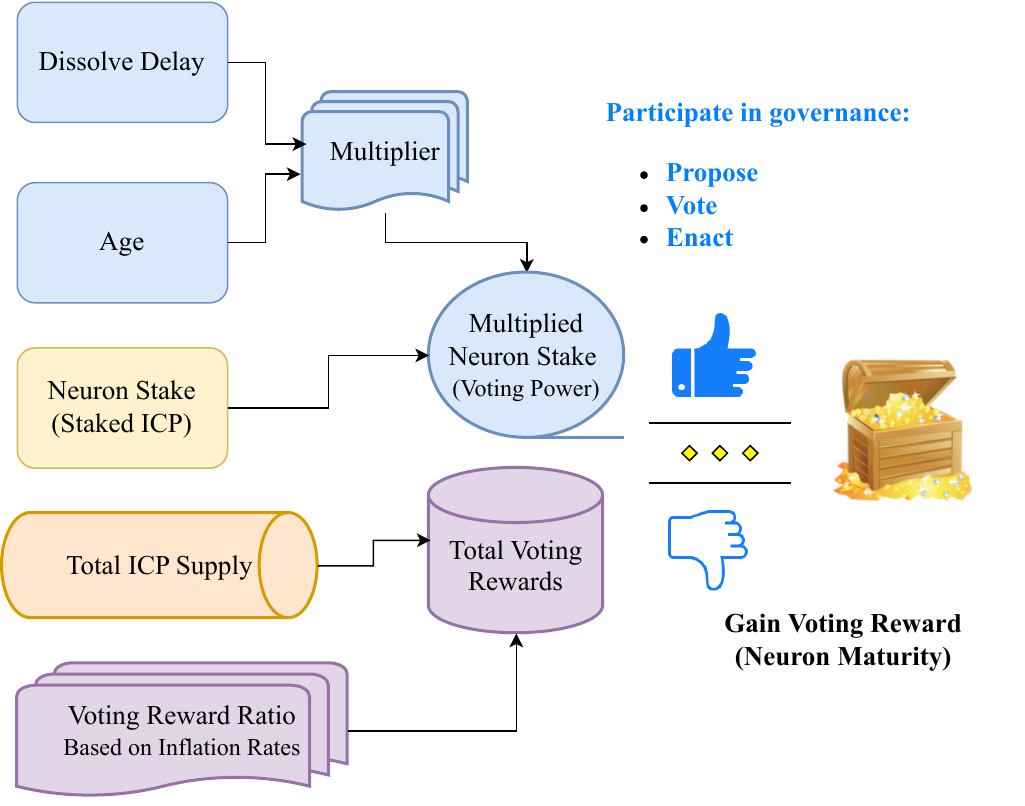}
    \caption{Network Nervous System (NNS) on the Internet Computer Protocol (ICP).}
    \label{fig: nns}
\end{figure}

%% file: figs/fig2.tex
\begin{figure}[!htbp]
\centering
\captionsetup{justification=centering}
\includegraphics[width=0.49\textwidth]{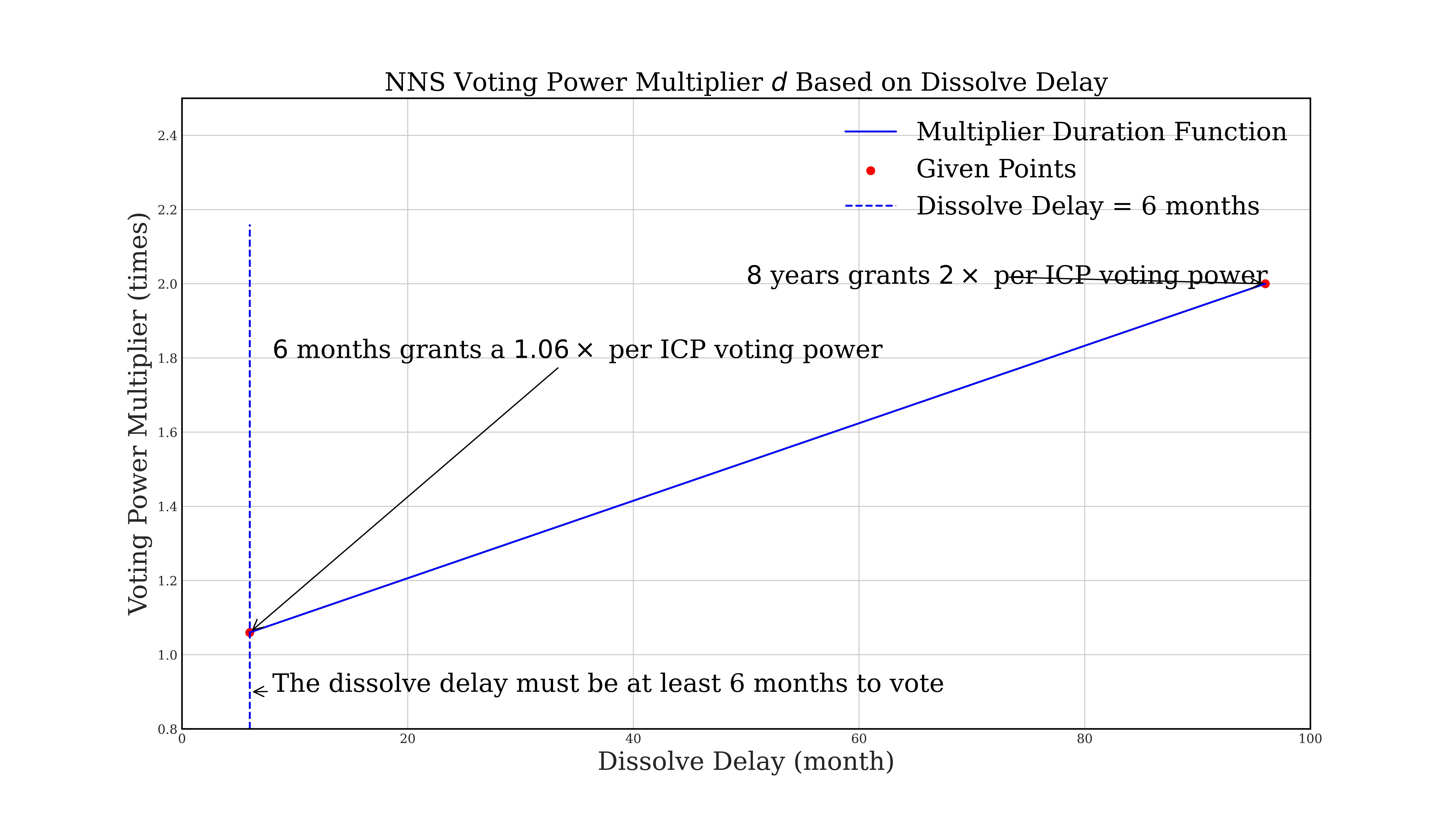}
    \caption{NNS Dissolve Delay Multiplier.}
    \label{fig: multiplier_duration}
\end{figure}

%% file: figs/fig3.tex
\begin{figure}[!htbp]
\centering
\captionsetup{justification=centering}
\includegraphics[width=0.49\textwidth]{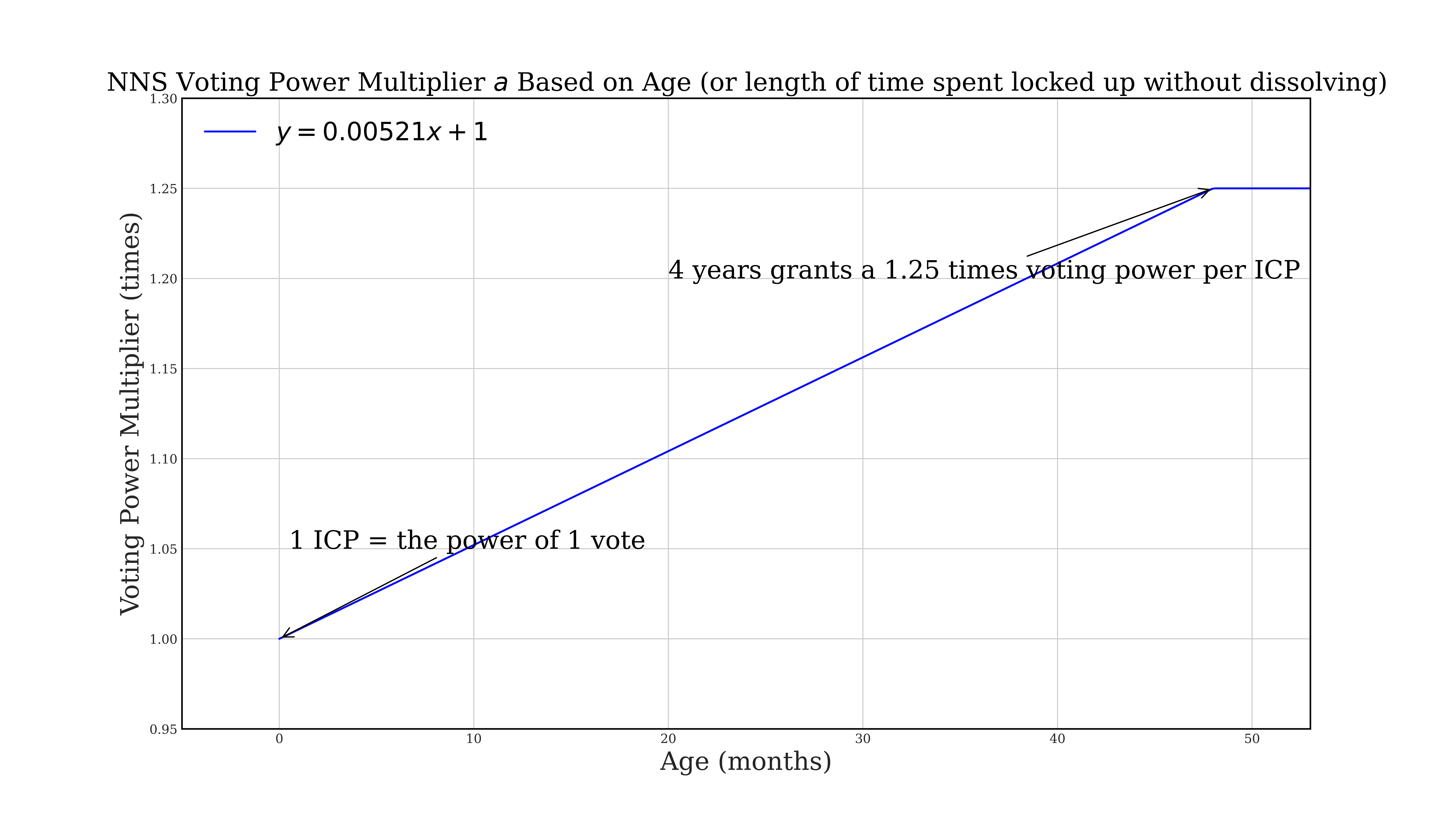}
    \caption{NNS Voting Power Age Multiplier.}
    \label{fig: multiplier_age}
\end{figure}

%% file: figs/fig4.tex
\begin{figure}[htbp!]
    \centering
    \includegraphics[width=0.49\textwidth]{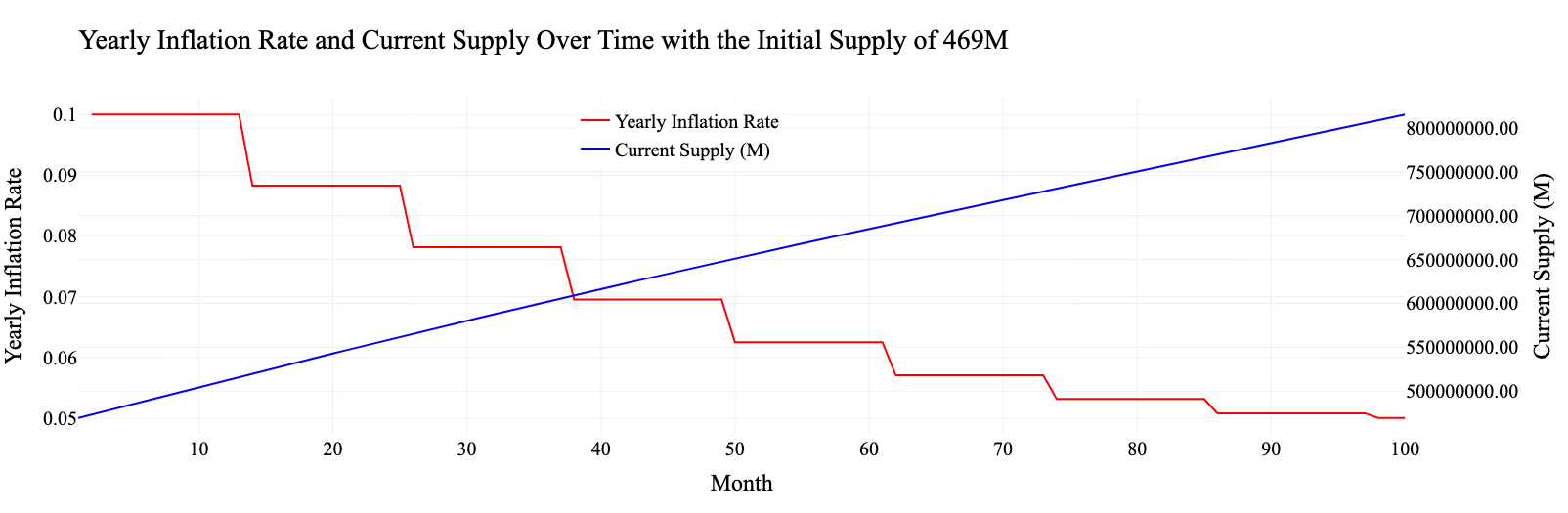}
    \caption{NNS Yearly Inflation Rate and Total Supplies}
    \label{fig:inflation}
\end{figure}

%% file: figs/fig5.tex
\begin{figure}[!htbp]
    \centering
    \includegraphics[width=0.49\textwidth]{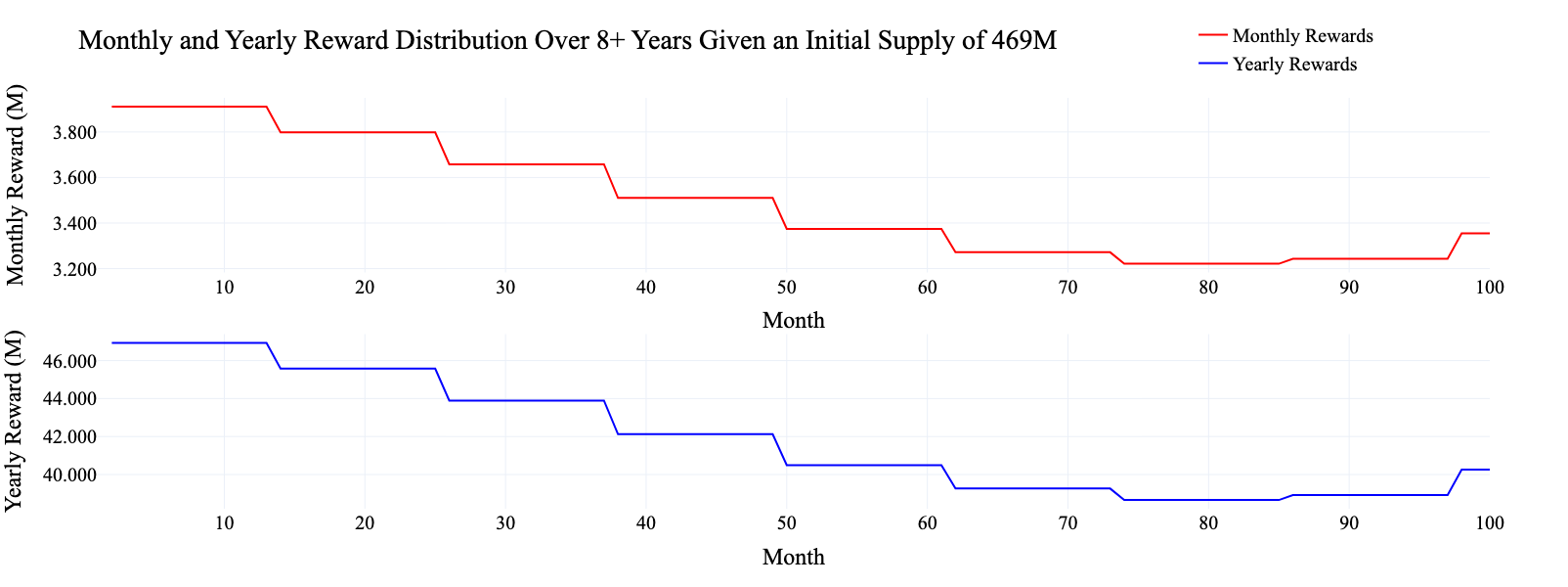}
    \caption{The total ICP monthly and yearly reward distribution for $8+$ years given the initial supply of $469M$}
    \label{fig:ICP_reward}
\end{figure}

%% file: figs/fig6.tex
\begin{figure}[!htbp]
    \centering
    \includegraphics[width=0.49\textwidth]{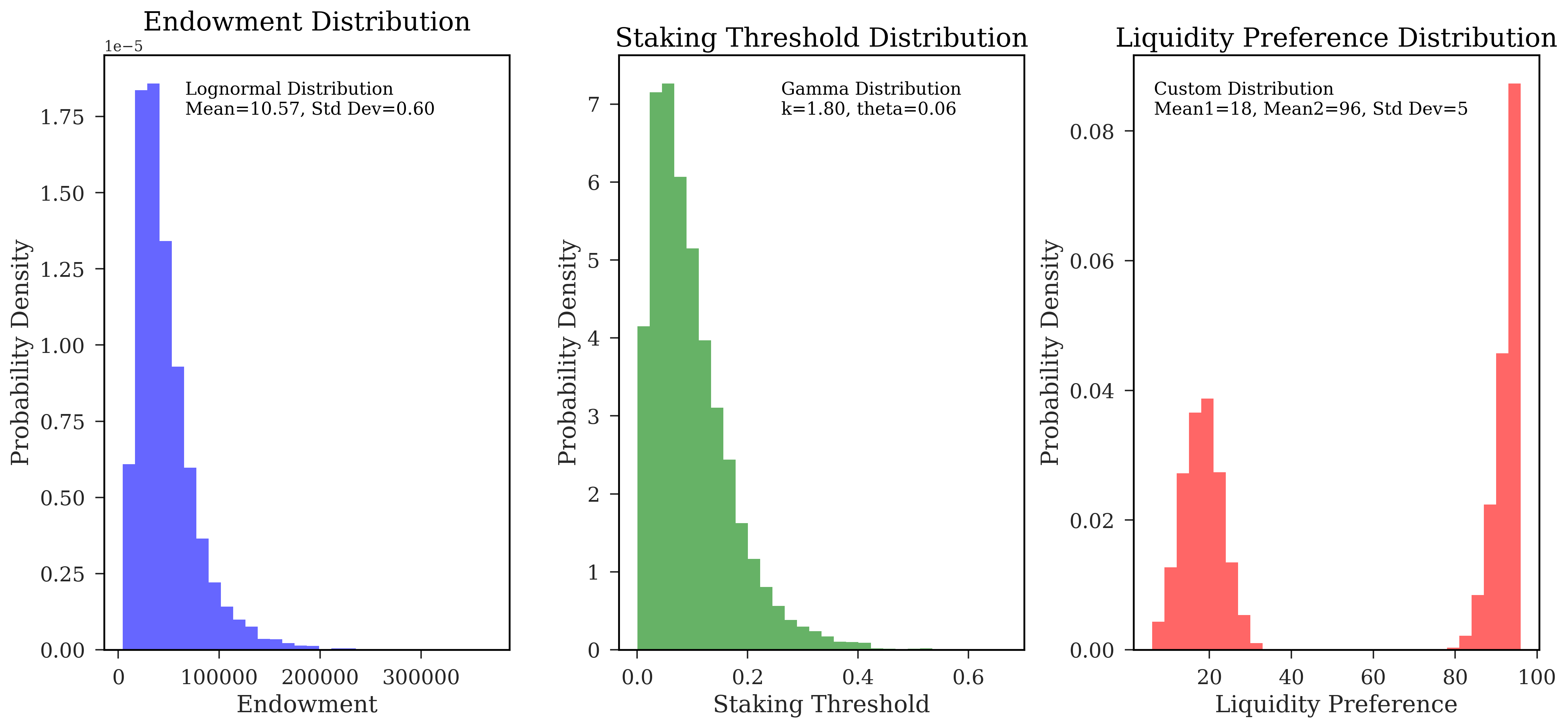}
    \caption{The distribution of agent endowment, staking threshold, and liquidity preference given the specified parameter configuration.}
    \label{fig:agent_initialize}
\end{figure}

%% file: figs/fig8.tex
\begin{figure}
  \centering
  \begin{minipage}{0.49\textwidth}
    \includegraphics[width=\linewidth]{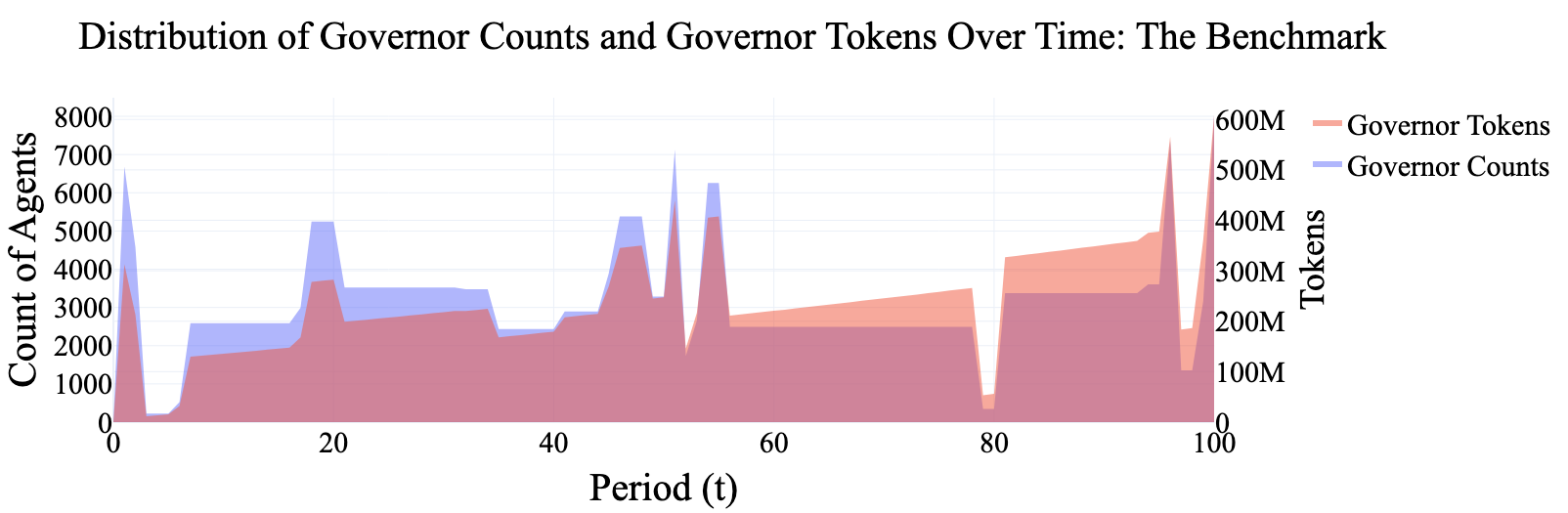}
    \label{fig:gov_benchmark}
  \end{minipage}
  \hfill
  \begin{minipage}{0.49\textwidth}
    \includegraphics[width=\linewidth]{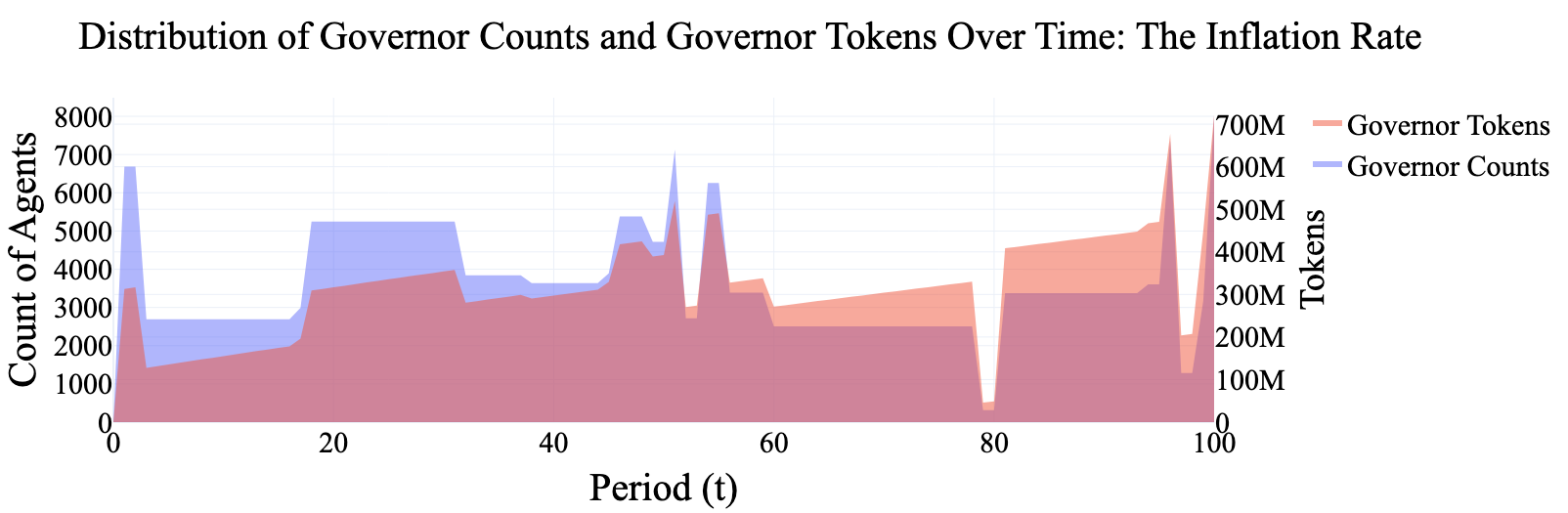}
    \label{fig:gov_inflation}
  \end{minipage}
  
  \begin{minipage}{0.49\textwidth}
    \includegraphics[width=\linewidth]{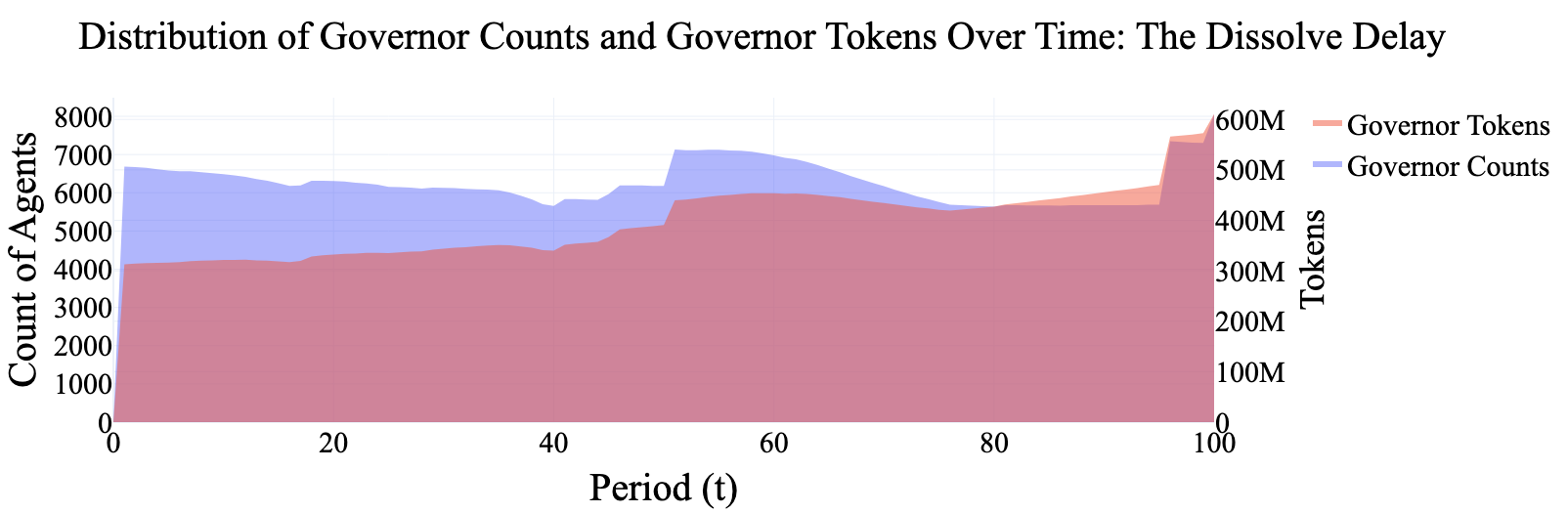}
    \label{fig:gov_dissolve_delay}
  \end{minipage}
  \hfill
  \begin{minipage}{0.49\textwidth}
    \includegraphics[width=\linewidth]{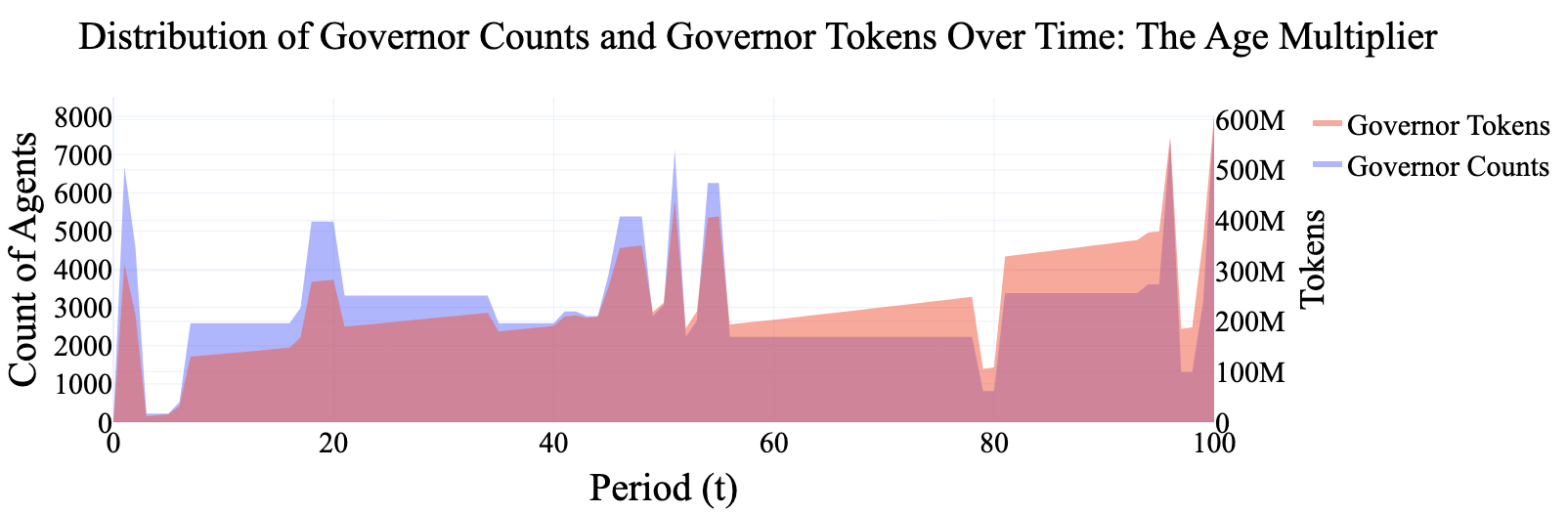}
    \label{fig:gov_age}
  \end{minipage}

  \begin{minipage}{0.49\textwidth}
    \centering
    \includegraphics[width=\linewidth]{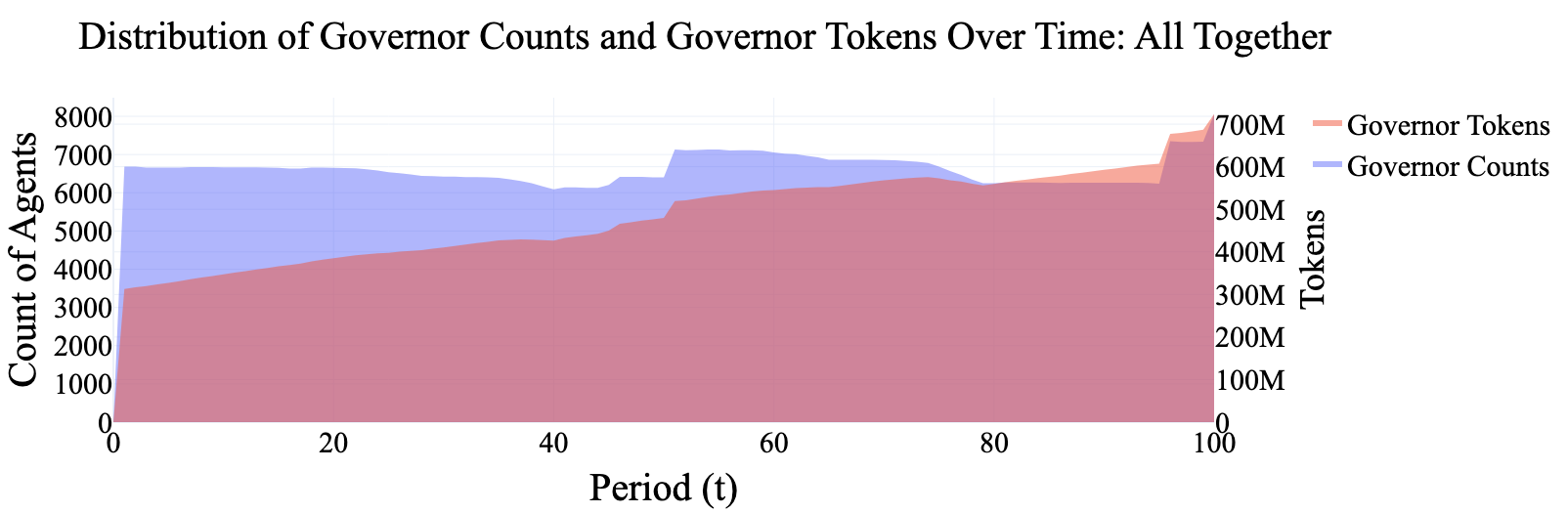}
    \label{fig:gov_all}
  \end{minipage}

  \caption{Comparative Studies on Governor and Governance Token Counts }
  \label{fig:gov_figures}
\end{figure}

%% file: figs/fig10.tex
\begin{figure}
  \centering
  \begin{minipage}{0.49\textwidth}
    \includegraphics[width=\linewidth]{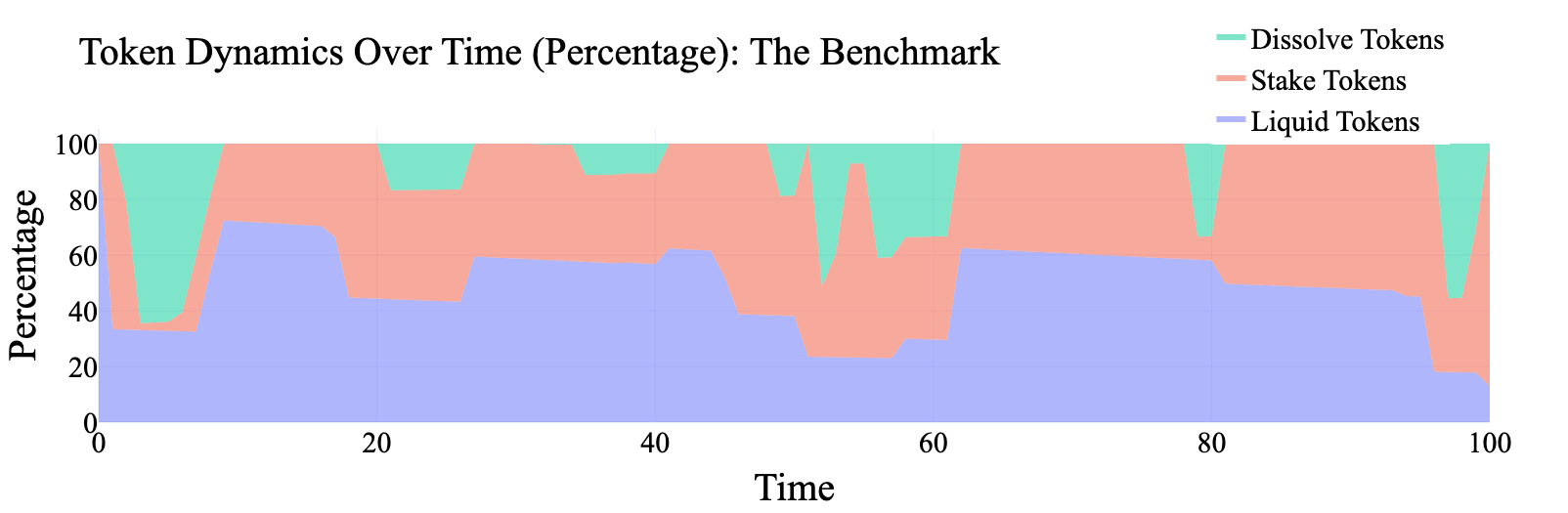}
    \label{fig:token_pct_benchmark}
  \end{minipage}
  \hfill
  \begin{minipage}{0.49\textwidth}
    \includegraphics[width=\linewidth]{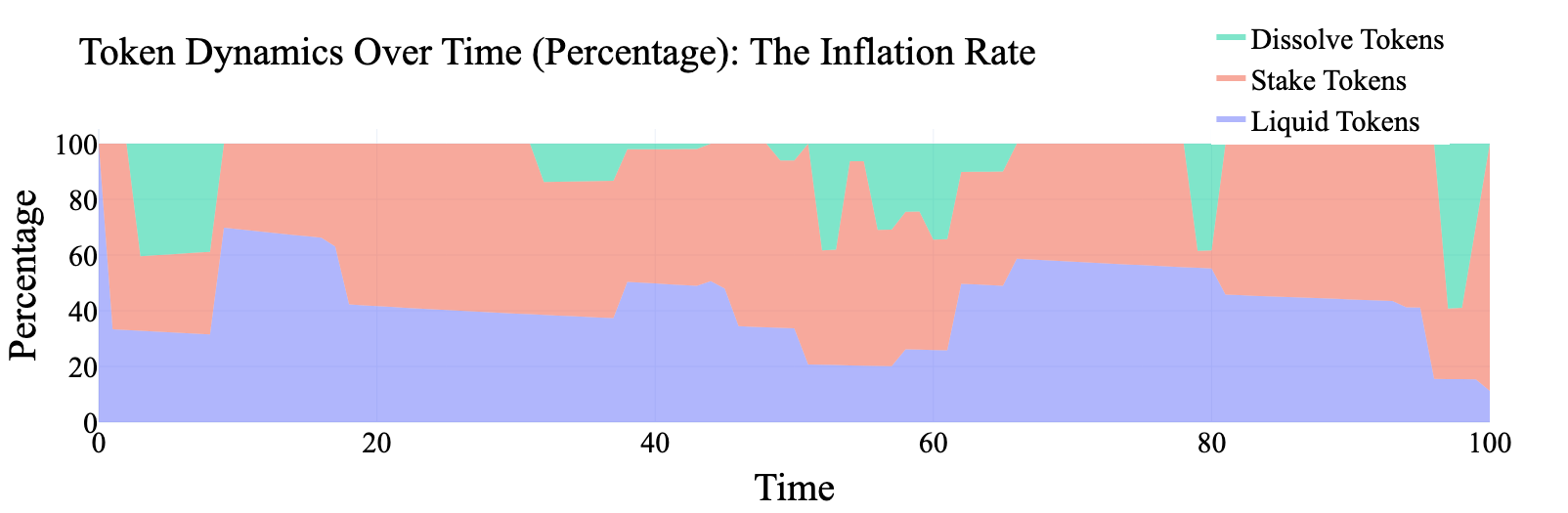}
    \label{fig:token_pct_inflation}
  \end{minipage}
  
  \begin{minipage}{0.49\textwidth}
    \includegraphics[width=\linewidth]{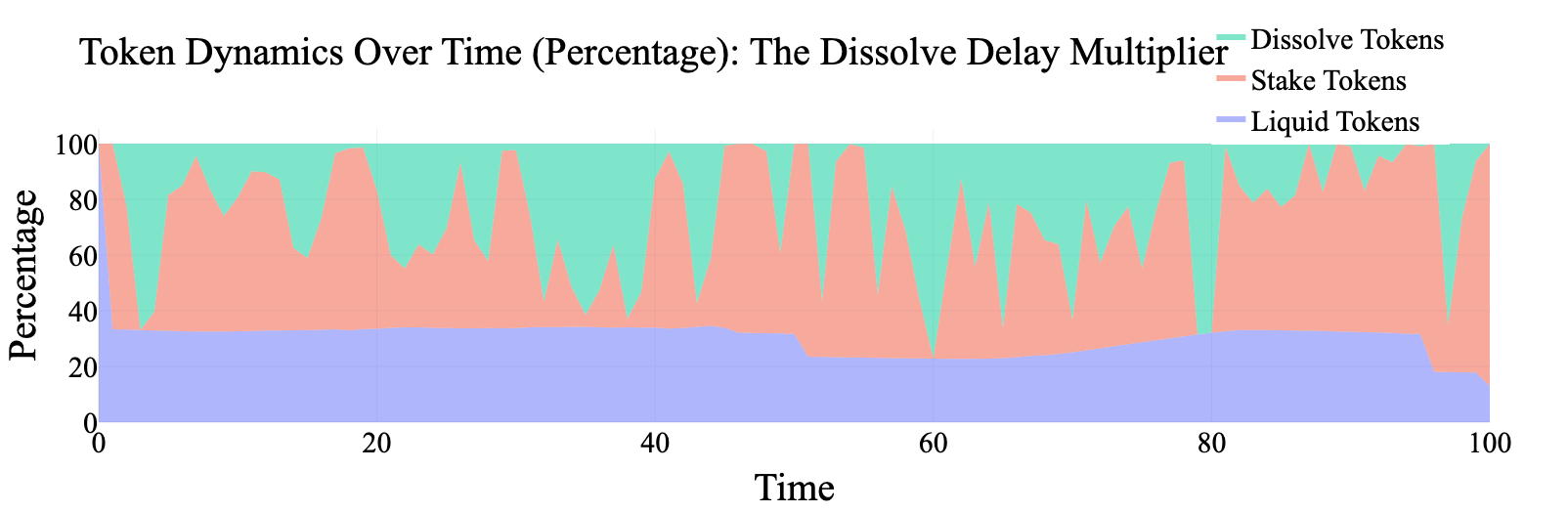}
    \label{fig:token_pct_dissolve_delay}
  \end{minipage}
  \hfill
  \begin{minipage}{0.49\textwidth}
    \includegraphics[width=\linewidth]{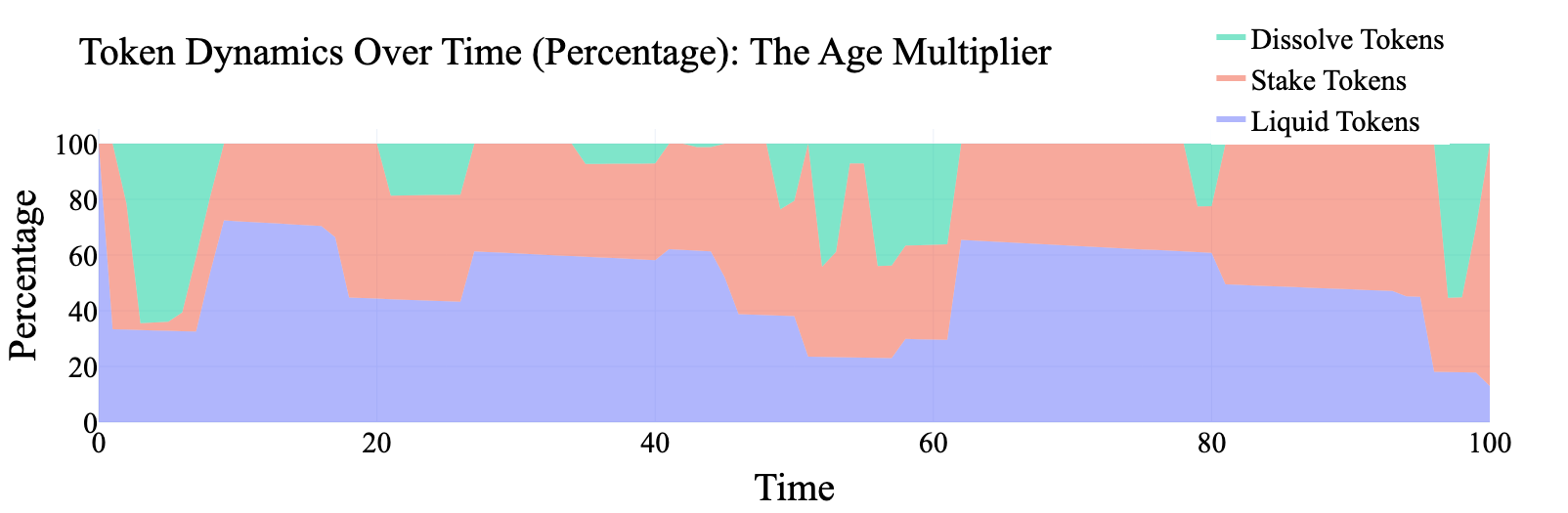}
    \label{fig:token_pct_age}
  \end{minipage}

  \begin{minipage}{0.49\textwidth}
    \centering
    \includegraphics[width=\linewidth]{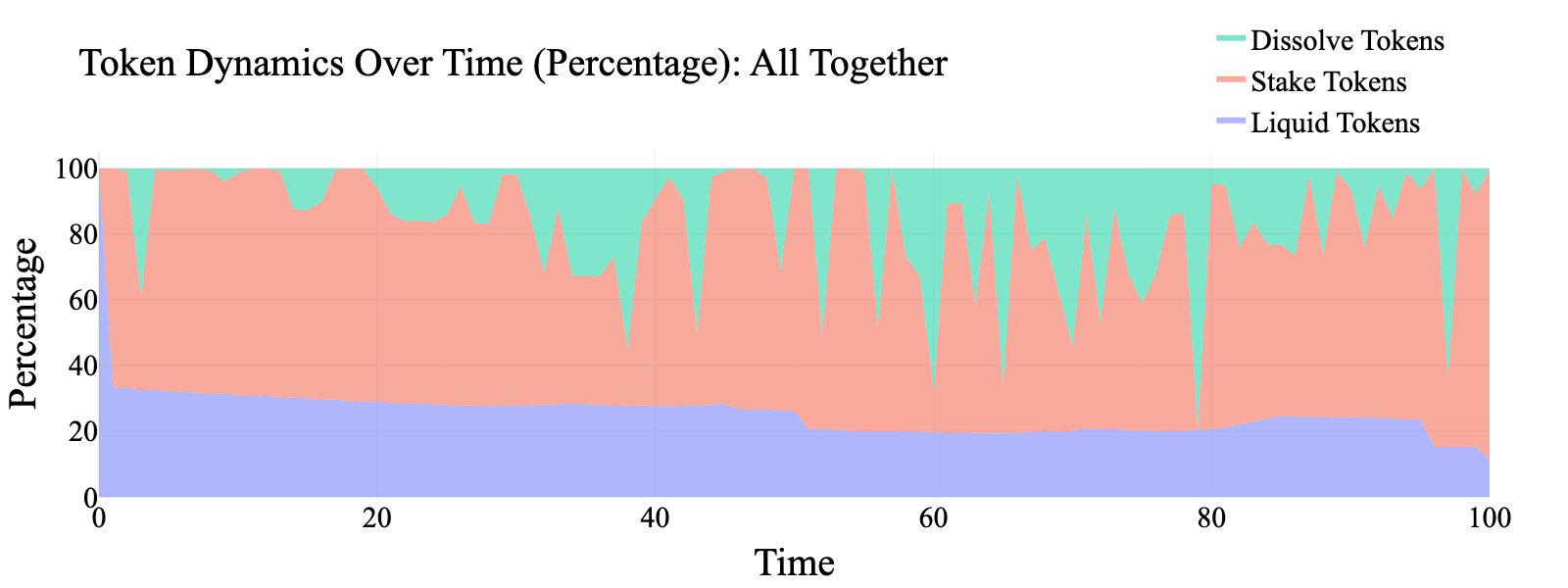}
    \label{fig:token_pct_all}
  \end{minipage}

  \caption{Comparative Studies on Token Distributions in Percentages}
  \label{fig:token_pct_figures}
\end{figure}